# A Framework towards Quantifying Human Restorativeness in Virtual Built Environments

## Zhengbo Zou,[1] and Semiha Ergan.[2]

[1]PhD Candidate, Department of Civil and Urban Engineering, New York University, 15 MetroTech Center, Brooklyn, NY, 11201; email: zz1658@nyu.edu
[2]Assistant Professor, Department of Civil and Urban Engineering, New York University, 15 MetroTech Center, Brooklyn, NY, 11201; email: semiha@nyu.edu

**ABSTRACT**

The impact of built environment on the human restorativeness has long been argued; however, the interrelations between neuroscience and the built environment, and the degree to which the built environment contributes to increased human restorativeness has not been completely understood yet. Understanding the interrelations between neuroscience and the built environment is critical as 90% of time in a typical day is spent indoors and architectural features impact the productivity, health and comfort of occupants. The goal of this study is to bring a structured understanding of architecture and neuroscience interactions in designed facilities and quantification of the impact of design on human experience. The authors first built two virtual environments (i.e., restorative and non-restorative) using the architectural designs features related to human restorativeness identified by previous research efforts. Next, user experiments were conducted in the two built virtual environments including 22 people. The subjects were asked to conduct navigational tasks while their bodily responses recorded by body area sensors (e.g., EEG, GSR, and Eye-tracking). The result showed that human responses in restorative and non-restorative environment had statistically significant difference. This study serves as the first step of understanding human responses in the virtual environment, and designing spaces that maximize human experience.

**INTRODUCTION**

Architectural design is essential to the health and well-being of residents, given the fact that we spend most of our times indoors (Hancock 2002). Various studies that investigated the impact of architectural design on human experience provided evidences that well-designed facilities can result in faster recovery in hospitals (up to 30%), better learning in schools (up to 25%), and higher productivity in offices (up to 25%) under variant configurations of architectural design features (e.g., natural daylighting, windows, exposure to nature) (Dudek 2000; Goldhagen, 2017). On the other hand, poorly designed buildings can have negative influence on the physical and psychological well-being of the residents. For example, Sick Building Syndrome (SBS), where the building occupants experience physical sickness and discomfort that appear to be linked to their time spent in the buildings (Dravigne et al., 2008). As a result, it is essential for researchers and practitioners in the Architectural Engineering and Construction (AEC) industry to understand if certain design features (e.g., presence/size of windows, level of natural light) have effect on human experience in buildings, and to have quantitative methods for measuring such impact.

Previous research studies investigating architectural design features' influence on human experience can be divided into two categories - subjective studies and objective studies. Subjective studies usually involve post construction surveys, where the researchers issue questionnaires to collect the opinions of the building residents towards the building design after the construction is complete, and the occupants had already been in the space for a period of time (Hedge et al., 1989). With the development of Virtual Reality (VR) technology, researchers started to use VR to create interactive and immersive virtual environments during the design phase to get future residents involved early on, with the goal of maximizing the residents' comfort once the construction is done (Du et al, 2017; Du et al., 2018; Du el al, 2018, Kasireddy et al., 2016). However, even with the VR technology, these studies still measure the satisfaction level of the residents by administrating surveys at the end of the VR study sessions. Admittedly, by surveying the residents after construction or during the design phase using VR can provide subjective feedbacks regarding the design directly from the building users to the architects. However, these survey studies are often time consuming and prone to errors caused by the selection of the sampling pool (i.e., resident demographics) and the subjective feelings of the surveyed.

In recent years, with the development of Body Area Sensor Network (BSN) technology and the improvement in motion capturing sensors, researchers started to look into the possibility of using BSN in combination with motion capturing systems to objectively and quantitatively study the human responses towards variant designed features by investigating the physical responses (e.g., heart rate, facial expression, and skin conductance) of the subjects when they are interacting with the virtually designed spaces (Radwan et al., 2017, Ergan et al., 2018, Zou and Ergan 2019). The results showed that it is possible to observe statistically significant differences among distinctly configured designed spaces, with stress and anxiety as the target human experience, when studying the subjects' BSN data. However, the causal relationship between design features and human experience in designed spaces is still inconclusive, since the differences detected in the body area sensors' data cannot be proved to be caused directly by the change of configuration of design features (Bratman et al., 2012).

Hence, there is still a need to pinpoint the causes of the variance in human experience in distinctly designed spaces, and eye tracking technology can be used to meet the need. Eye-trackers are devices that measure the eye-movements of the users, which can be used to reflect the users' gaze pattern and visual attention (Dzeng et al., 2016). Human eye-movement and gaze pattern have been linked to human attention and cognition, which reveal the brain activities (e.g., scene understanding and danger awareness) of the users (Hasanzadeh et al., 2016). Therefore, by measuring the eye-movement, one could pinpoint the visual attention of the users, and in turn understand the causal relationship between the visual stimuli (e.g., design features) and the human response.

This paper proposed a framework to quantify the human restorativeness in virtual environments configured with distinct architectural design features. The authors conducted user studies in the virtual environment and collected quantitative data using BSN sensors. During the user study sessions, the subjects were directed to conduct navigate (e.g., to find a certain room) in the virtual environments. While they were navigating in the VEs, they were exposed to variant design features (i.e., presence/size of windows, level of natural light). The design features were embedded in the virtual environments in a polarizing way. In other words, the virtual environments were defined as either restorative or non-restorative based on the parameters of the features. For example, in the restorative environment, the room can be enclosed by glass curtain walls and have ample natural light, while in the non-restorative environment, there is only closed walls and artificial light. Furthermore, this paper provided a data analysis method for analyzing BSN data collected during user studies, when the subjects were interacting with virtual environments containing distinctly design features. The data analysis method used statistical tests (e.g., paired t-test for EEG data) to find if the BSN data is statistical significantly different in distinctly designed spaces. The objective of this method is to provide a generic data analysis algorithm that can be used to make sense of the BSN data in order to help architects to quantitatively understand the impact of design choices on human experience.

**BACKGROUND**

This paper builds on the existing body of knowledge from the following domain: (1) Architectural design features' impact on human restorativeness in buildings; and (2) BSN metrics and data analysis techniques.

**Architectural Design Features' Impact on Human Experience in Buildings**

Architectural design features have a wide range of influence on human experience in buildings (Eberhard 2009, Zou and Ergan 2019). Sense of restorativeness was studied in the architectural domain since it is closely related to the functionality of a variety of buildings, such as hospitals, schools and office buildings (Beukeboom et al., 2012; Parsons et al., 1998; Rashid and Zimring 2008). The sense of psychological restoration is an indicator of human well-being and satisfaction and implies the recovery from stress or attentional fatigue (Kaplan 1995). A restorative environment has the architectural design qualities that reduce cognitive fatigue and stress (Evans 1998). The authors' previous crowdsourcing study including 400 participants and expert architects identified a range of design features that are linked to the sense of restorativeness, including presence of windows, size of windows, presence of nature view and presence of natural light. The architectural design features' impact on the sense of restorativeness are summarized in Table 1.

**Table 1. Mapping of architectural design features and human restorativeness**

| Design Features | Impact on Human Experience |
| --- | --- |
| Presence of windows | Presence of windows increase the speed of recovery from stress and attention fatigue. |
| Size of windows | Small windows prevent occupants from restoring from fatigue. |

| Presence of natural light | Natural light helps relax and ease the occupants' stress level. |
| Presence of nature view | Nature view induces refreshing and restorative feelings. |

**BSN Metrics and Data Analysis Techniques**

Various metrics have been proposed to understand the implications of visual stimuli and architectural layouts on emotional and physiological states of human beings (Parsons et al. 1998). Commonly used metrics include heart rate, blood pressure, and startle reflex to measure human emotional experience such as stress, pain, anxiety and memory (Parsons et al. 1998). Biometric sensors provide a great opportunity for gaining deeper insights into human experience in the environment. It also provides real-time unbiased data on physiological responses. A body area sensor network (BSN), representing a collection of biometric sensors, is a set of biometric wearable sensing devices connected to a system through wireless networks (Ullah et al. 2012). BSN enabled researchers to measure and monitor such metrics and to relate those metrics with the events and stimuli presented to the subjects. Data collected from such technologies helps to relate bodily measures to human experience and quantify this relationship.

While capturing bodily measures, biometric sensors help in removing the bias in the data received through the traditional methods (i.e., self-reports, observations). It provides an opportunity for verification of results by collecting quantitative measurements besides the qualitative data perceived from the traditional methods to offer valid and reliable results. In order to achieve the maximum flexibility with the study setup, only non-invasive BSN (i.e., with electrodes that are placed on the skin or scalp) were used in this study to capture a real-time cognitive, emotional, and physiological characteristic of people as they interact with different configurations of virtual architectural spaces. Table 2 provides an overview of sensors used and the metrics investigated in this study.

**Table 2. Sensor measurements and definitions of metrics used in the analysis**

| Sensor | Analysis metrics | Definition of metrics |
|--------|------------------|------------------------|
| EEG | Power Spectrum | A power spectrum diagram shows EEG signals in the frequency domain and describes the distribution of power in frequency bands. |

| | Frequency Bands | Representation of EEG oscillations in the frequency domain instead of the time domain. Includes Delta (0-4 Hz), Theta (4-8 Hz), Alpha (8-14 Hz), Beta (14-40 Hz), and Gamma (40 Hz and above) bands. |
|---|---|---|
| GSR | SCR peaks | SCR signal jump higher than the defined threshold (i.e., onset amplitude). |
| | Peak Amplitude | Amplitude at the peak minus the defined threshold (i.e., onset amplitude). |
| Eye-tracker | Area of Interest (AOI) | Video frames containing the research interests (e.g., design features). |
| | Heatmap | Distribution of visual attention. More hot spots mean more distracted attention. |
| | Number of fixations | Number of periods when the respondents' eyes are locked toward a specific AOI. A large number of fixation indicates complex tasks and the searching and navigating efficiency is low. |
| | Time to First Fixation (TTFF) | The amount of time the respondents spent to look at a specific AOI for the first time. Smaller TTFF indicates high attention level and a more focused mind. |
| | Time spent | The amount of time a respondent spent on an AOI. Less time spent indicates high attention level. |

As summarized in Table 2, electroencephalogram (EEG) is an electrophysiological monitoring tool to record electrical activity of the brain by placing electrodes along the scalp. EEG helps in obtaining insights into how human brain works and reacts towards different spatial settings. The benefits of using EEG come from the fact that it is non-invasive and can provide high quality and good time resolution of brain activity (Clemente et al. 2014; Pham and Tran 2012). It's also considered as one of the most intensive biometric research tools since it provides data for both emotional valence and arousal. Several emotional statuses such as meditation, stress, and engagement are accompanied by distinguishable electrical brain activation patterns that can be identified by EEG. Besides brain activity, galvanic skin response (GSR) is a powerful tool to measure a widely referred metric called skin conductance. GSR provides data on the amount of sweat secretion from skin, and is reported to have a positive correlation with the magnitude of emotional arousal (Villarejo et al. 2012) as shown in Table 2.

Eye-tracking metrics can be divided into two main groups, fixations and saccades. A fixation is a relatively motionless gaze of an Area of Interest (AOI), which is defined as specific parts of a product or a design the researchers interested in. On the other hand, a saccade is defined as the fast movement between two fixations (Jaimes and Sebe 2007). Fixation related metrics include fixation location and duration. Fixation location represents the subjects' attention and fixation duration indicates the level of cognitive difficulty (Aries et al., 2010). Saccades represent rapid eye-movements, and no visual information is recognized and understood by the user (Hasanzadeh et al., 2016), so only fixation related metrics were used in this study. A summary of the fixation related metrics is presented in Table 2.

Previous research studies using BSN focused on interpreting the results statistically. Researchers looked into simple statistics (e.g., mean, standard deviation, and min/max values), as well as statistical tests result (e.g., p-value from t-test). Simple statistics drew from the BSN metrics can be used to interpret the average behavior of subjects. Statistical tests were used to process BSN data when there were comparisons of multiple AOIs, or multiple groups of people.

In summary, given all these sensors available and what they measure, it is evident that a BSN configured to understand the impact architectural design features on human experience should include EEG, GSR and eye-tracking sensors, since those sensors are all non-invasive, compatible with a VR setup, requires a reasonable effort for placement and data analysis, and complement each other to get a good understanding of what a user is experiencing in an architectural setting.

**EXPERIMENT SET-UP AND USER STUDIES**

The authors conducted user studies to investigate the architectural design features' impact on human experience using BSN. There were 22 participants recruited for the study, all of whom were graduate student with previous knowledge of architectural design and construction. There were thirteen male participants and nine females. Two polarizing virtual environments, which were marked as restorative and non-restorative, were created for the user studies. The subjects were asked to navigate in the virtual environments during the study. Their bodily response was collected using EEG, GSR and eye-tracker. The data analysis method used the BSN metrics as inputs and conducted paired t-test for all participants from restorative and non-restorative environments to find if the design features could cause differences in the subjects' response between two environments.

The virtual environments were first modeled based on a real academic building in New York City using a BIM authoring tool, and then converted to interactive virtual models using a game engine. The two environments are identical except for certain design feature changes, as shown in Table 3. There were four architectural design features implemented in the virtual environments (i.e., presence of windows, size of windows, presence of nature view, and presence of natural light), because of the implication of influence these features have on human restorativeness. To elaborate, the restorative environment was designed to have more windows and larger size windows in the corridors and rooms than the non-restorative one. While the restorative environment has ample natural light and a view of nature, the non-restorative one only has artificial lights and no nature view. The details of the features implemented in the user studies are shown in Table 3.

**Table 3. Architectural design features implemented in the virtual environments**

| Design Features | Restorative VE | Non- Restorative VE |
| --- | --- | --- |
| Presence of windows | Has windows in the measured corridor. | No windows in the measured corridor. |
| Size of windows | Large windows in the measured room (60% windows to wall ratio). | Small windows in the measured room (20% windows to wall ratio). |
| Presence of nature view | Connection to external environments and a view of natural plants. | No connection to external environments and no nature view. |
| Presence of natural light | Ample natural light (300 lx) because | No natural light, only artificial lights |

of windows in the measured corridor.     in the measured corridor.

During the user studies, 22 participants were asked to navigate in two virtual environments. The subjects were directed to memorize the current and desired room temperature so that they can change it in the mechanical room in the virtual environments. While the subjects were navigating in the VEs, various design features (i.e., features from Table 3) were implemented in the corridors and rooms where the subjects were supposed to "walk-through" in the virtual environments. Screenshots of the design features "presence of windows" and "presence of nature view" are shown in Figure 1a and Figure 1b. The subjects' BSN metric measurements were collected while they were interacting with the virtual environments.

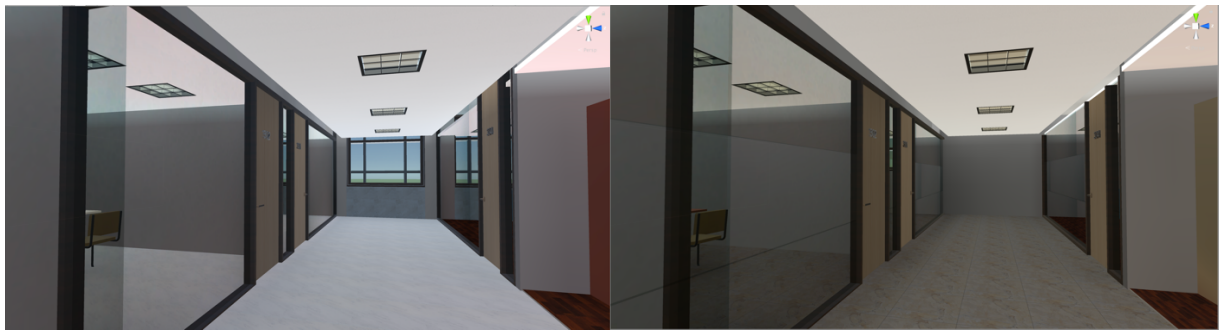

**Figure 1a. Presence of windows positive (left) and negative (right) environments**

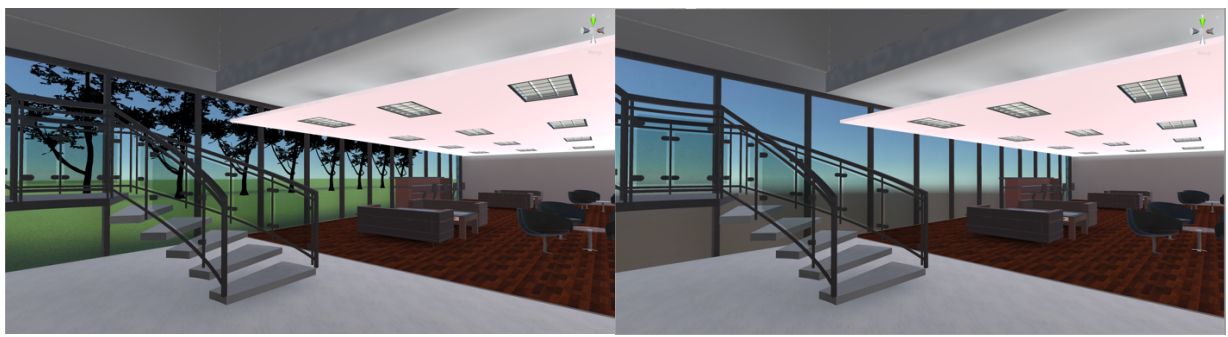

**Figure 1b. Presence of nature view positive (left) and negative (right) environments**

A widely used wireless mobile EEG device, a wireless GSR device and a mobile eye-tracking device were used in the user studies, as shown in Figure 2a. The virtual environments were presented on a 98-inch screen equipped with motion capturing cameras. The head-movement of the subjects were tracked using the cameras, and the subjects' field of view changed in the virtual environments accordingly. For user interaction in the virtual environments, the subjects used an apex device, which has a joystick on the top, and a trigger button at the bottom. Figure 2b shows a subject when she was conducting the experiments.

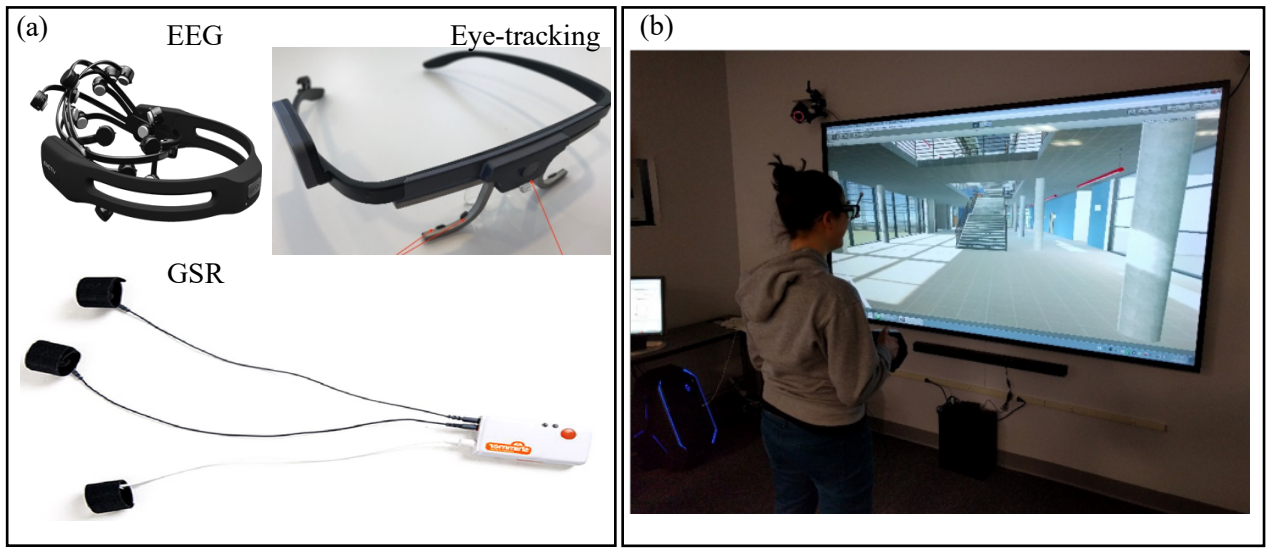

**Figure 2a (left) EEG, GSR, and Eye-tracker used in user studies; Figure 2b (right) Subject navigating in virtual environments.**

The user study was designed to last around 40 minutes in total. The break-down of the time allocated during the user study is shown in Figure 3. The subjects had a short training to get familiar with the equipment and the user interaction methods (using joystick for moving and motion tracking camera for head movement) in the virtual environments. Following that, the subjects started the experiment with a randomly chosen restorative or non-restorative environment, where they navigated in the environments. Then the subjects would be presented the other virtual environment.

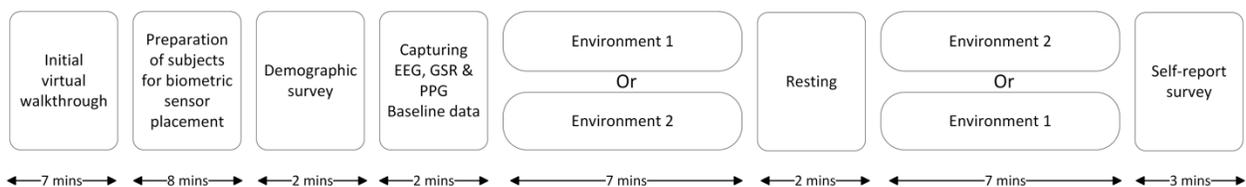

**Figure 3. User study process and approximate time**

## RESULTS AND DISCUSSION

The authors analyzed each sensor data individually, and statistical analysis was used to draw conclusions of the statistical significance between the architectural design features and the human experience. Additionally, eye-tracking data was analyzed to find the causal relationship by looking into the subjects' attention.

**EEG Data Analysis**

The EEG data was analyzed using power spectrum analysis by analyzing individual signal activity on each channel of each respondent on each frequency band. The EEG headset incorporates 14 channels and positioned on the international 10-20 system, with a sampling rate of 128 Hz. The raw EEG signal from all channels is analyzed to calculate and plot power spectrum diagrams to quantify the differences in EEG signals in each virtual environment. A power spectrum diagram describes the distribution of power in frequency bands, which are Delta (0-4 Hz), Theta (4-8 Hz), Alpha (8-14 Hz), Beta (14-40 Hz), and Gamma (40 Hz and above). To generate power spectrum diagrams, first, the raw EEG signal of each participant was normalized against their baseline signals. For a given participant's EEG data, this normalization includes calculating the mean value of EEG signals collected during the baseline period, and dividing the EEG signals captured beyond the baseline period by the mean value of the baseline period. The goal of this normalization is to bring different subjects' EEG data to a comparable scale.

Next, Fast Fourier Transformation (FFT) was conducted using a moving window of one second to convert the EEG signal from the original time domain to the frequency domain. The signal was then filtered to remove any noise using a high pass filter of 0.5 Hz and a low pass filter of 50 Hz in order to calculate the power of each channel, and hence the power of the different frequency bands. For each participant, the average normalized power across time was calculated for the EEG signals on each channel in each frequency band.

When the power spectrum diagrams are analyzed across participants, it was apparent that there were statistically significant differences in oscillations on each channel when participants were in VE1 and VE2. Paired sample t-test at 95% confidence interval compared the mean responses of subjects for each channel between the two VEs. The results conducted for each participant separately on each channel across frequency bands (i.e., Delta, Theta, Alpha, Beta, and Gamma) showed significant difference between two environments as shown in Table 4. The percentage is calculated by using the number of participants that showed significant difference in EEG signal data divided by the total number of participants.

**Table 4. Percentage of participants that showed statistically significant difference in EEG signal data in VE1 and VE2 for each channel across all frequency bands**

| Channels | Delta | Theta | Alpha | Beta | Gamma |
|----------|-------|-------|-------|------|-------|
| F7 | 73% | 98% | 93% | 97% | 97% |
| F8 | 76% | 77% | 93% | 99% | 93% |
| F3 | 70% | 83% | 87% | 97% | 87% |
| F4 | 90% | 83% | 96% | 93% | 95% |
| FC5 | 87% | 80% | 90% | 87% | 97% |
| FC6 | 78% | 87% | 95% | 93% | 97% |
| AF3 | 77% | 83% | 93% | 97% | 98% |
| AF4 | 81% | 87% | 93% | 97% | 90% |
| T7 | 84% | 83% | 97% | 90% | 88% |
| T8 | 80% | 83% | 90% | 89% | 96% |
| P7 | 77% | 83% | 97% | 98% | 90% |
| P8 | 87% | 87% | 87% | 94% | 93% |
| O1 | 80% | 75% | 91% | 92% | 97% |
| O2 | 63% | 93% | 93% | 92% | 93% |

Results of raw data analysis on each channel across participants (Table 4) show that a high number of people (more than 70% up to 100%) showed oscillations in Alpha, Theta, Gamma, and Beta bands across frontal channels that are statistically significantly different between the two VEs. This result was the same for the frontal channels for all frequency bands. In conclusion, the result concurs with the hypothesis that the subjects experience statistically different EEG response between restorative and non-restorative virtual environments.

**GSR Data Analysis**

A GSR sensor was used in this study to measure the subject' skin conductance in units of micro-Siemens (µS). The sampling frequency was set to 104 Hz. Disposable Ag/AgCI electrodes were placed on the palm side of the index finger and the middle finger of the non-dominant hand of the subject to avoid any noise from moving the hand in the task.

The data was analyzed by determining the number of peaks in the skin conductance response (SCR) that subjects had during the experiment based on the following logic. The phasic data (SCR), which corresponds to faster-varying skin conductance data component (fluctuating within seconds), was extracted from the GSR signal by using an averaging sliding window. The averaging sliding window computes the median GSR for 4000 milliseconds (ms) that is centered on a current sample. This value is then subtracted from the current sample to get the phasic data. Peaks onset/offset thresholds were set to 0.01 µS/0 µS (Benedek and Kaernbach, 2010). Peak onset value represents the starting point in time where a peak is detected, while the offset value represents the time when a peak has passed. To avoid false positives, the onset value is not counted if it is less than 0.01 µS. The maximum original GSR data within each pair of onsets and offsets is a SCR peak. GSR peak amplitude is the amplitude at the peak minus the amplitude at onset. A peak is only considered if its amplitude was higher than the threshold amplitude by 0.005 above the onset value. Also, a signal jump threshold that accounts for false peaks - caused by noise - is set to 0.02 µS (iMotions 2017).

Table 5 shows the average number of the SCR peaks as well as the peaks amplitude of all subjects during the baseline period, VE1, and VE2. As expected, it can be seen that the average number of SCR peaks as well as the amplitude of the peaks increased from the baseline to VE1 and to VE2. These results, when coupled with EEG analysis, can be interpreted for increased anxiousness (higher emotional response) with respect to baseline, with more impact in the negative environment (Villarejo et al. 2012).

**Table 5. Analysis of GSR data across participants**

|                        | Baseline | Restorative VE | Non-Restorative VE |
|------------------------|----------|----------------|--------------------|
| **Average # of SCR peaks** | 2.54     | 6.73           | 12.35              |

| Average # of peaks/min | 5.01 | 10.62 | 21.22 |
|---|---|---|---|
| Peaks amplitude (µS) | 0.03 | 0.04 | 0.05 |

**Eye-tracking Data Analysis**

The authors also conducted data analysis for eye-tracking metrics. The results are averaged across 22 participants, and shown in Table 6. On average, the subjects showed less time spent in the positive environment for each AOI than the negative one, indicating the subjects were faster on the executing tasks in the positive environment. When the complexity of the tasks is identical, less time spent means higher attention level in the environment. The subjects also had a smaller number of fixations in the positive environment for all AOIs comparing to the negative one. A large number of fixations indicates the task is complex and the subjects showed decreased efficiency in searching and navigating tasks. Subjects showed smaller Time to First Fixation (TTFF) for the AOI: presence of window, level of natural light, and exposure to nature views. TTFF is calculated as the time in milliseconds that the subjects took to notice a specific AOI, and smaller TTFF indicates the high attention level and more focused mind.

**Table 6. Differences in the eye tracking metrics used for both VEs**

*: R: Restorative environment; NR: Non-restorative environment.

| AOI | Time Spent (seconds) | | Number of Fixations (times) | | Time to First Fixation (millisecs) | |
|---|---|---|---|---|---|---|
| | R* | NR* | R | NR | R | NR |
| AOI with Windows | 3.5 | 3.0 | 270 | 290 | 1.0 | 3.0 |
| AOI with different size of windows | 3.8 | 3.1 | 300 | 335 | 2.1 | 3.0 |
| AOI with higher natural daylight | 5.3 | 4.9 | 420 | 620 | 1.5 | 5.2 |
| AOI with exposure to nature (or restorative nature images) | 6.8 | 8.4 | 650 | 900 | 2.1 | 2.9 |

**CONCLUSION**

This study serves as one of the first studies that aimed at providing empirical quantification on human experience in architectural spaces using an integrated and rapidly replicated VR and BSN based data collection. The results provide evidence that the presence/size of windows, natural light, and natural views are indeed related to restorativeness experience (statistically significant) in designed spaces and can positively impact the residents if present. The experimental procedure proposed in this study can be used as a general experimental method for architects to use when collecting user experience data during the design process for improving occupants' experiences. Finally, with more participants, the results could serve as guidelines (i.e., how to correctly configure architectural features) when designing spaces with human restorativeness in mind.

**REFERECENS**